\documentclass[final]{llncs}
\usepackage[usenames,dvipsnames]{xcolor}
\usepackage{graphicx}
\usepackage{tabularx}
\usepackage{verbatim}
\usepackage{multirow}
\usepackage{microtype}
\usepackage{mathtools}
\usepackage{siunitx}
\usepackage{caption}
\usepackage{colortbl}
\usepackage{booktabs}
\usepackage{csquotes}

\colorlet{light-gray}{gray!20}

\newcommand{\acol}{{\sc Acol}}

\newcommand{\best}{\cellcolor{YellowGreen}}

\usepackage{fancybox}
\newenvironment{custom}%
{\vskip\baselineskip\VerbatimEnvironment
\scriptsize\begin{Sbox}\begin{BVerbatim}}
{\end{BVerbatim}%
\end{Sbox}\noindent\centerline{\TheSbox}\vskip\baselineskip}
\usepackage{hyperref}
\usepackage[capitalize]{cleveref}

\usepackage[misc]{ifsym}

\usepackage{xcolor}
\definecolor{orcidlogocol}{HTML}{A6CE39}
\usepackage{tikz}
\usetikzlibrary{svg.path}
\tikzset{
  orcidlogo/.pic={
    \fill[orcidlogocol] svg{M256,128c0,70.7-57.3,128-128,128C57.3,256,0,198.7,0,128C0,57.3,57.3,0,128,0C198.7,0,256,57.3,256,128z};
    \fill[white] svg{M86.3,186.2H70.9V79.1h15.4v48.4V186.2z}
                 svg{M108.9,79.1h41.6c39.6,0,57,28.3,57,53.6c0,27.5-21.5,53.6-56.8,53.6h-41.8V79.1z M124.3,172.4h24.5c34.9,0,42.9-26.5,42.9-39.7c0-21.5-13.7-39.7-43.7-39.7h-23.7V172.4z}
                 svg{M88.7,56.8c0,5.5-4.5,10.1-10.1,10.1c-5.6,0-10.1-4.6-10.1-10.1c0-5.6,4.5-10.1,10.1-10.1C84.2,46.7,88.7,51.3,88.7,56.8z};
  }
}
\renewcommand{\orcidID}[1]{%
  \resizebox{8px}{8px}{
      \href{https://orcid.org/#1}{\tikz[yscale=-1,transform shape]{\pic{orcidlogo}}}}%
}

\begin{document}
\pagestyle{empty}  % switches off printing of running heads

\title{On the Performance of\\ Bytecode Interpreters in Prolog}

\author{Philipp K\"orner $^\textrm{\Letter}$ \orcidID{0000-0001-7256-9560},
        David Schneider
        and Michael Leuschel \orcidID{0000-0002-4595-1518}}

\institute{Institut f\"{u}r Informatik, Heinrich Heine University D\"{u}sseldorf, Germany
 \\
  {\tt\scriptsize \{p.koerner, david.schneider, leuschel\}@hhu.de}
 }

\maketitle
\begin{abstract}
The semantics and the recursive execution model of Prolog make it very natural
to express language interpreters in form of AST (Abstract Syntax Tree) interpreters
where the execution follows the tree representation of a program.
An alternative implementation technique is that of bytecode interpreters.
These interpreters transform the program into a compact and linear representation
before evaluating it and are generally considered to be faster and to make
better use of resources.

In this paper, we discuss different ways to express the control flow of interpreters in Prolog
and present several implementations of AST and bytecode interpreters.
On a simple language designed for this purpose, we evaluate whether techniques
best known from imperative languages are applicable in Prolog and how well they perform.
%We designed a simple language that consists of integer assignments, basic arithmetic and logic operators, as well as if- and while-statements.
%Furthermore, we evaluate whether known techniques of interpreter design are applicable in Prolog.
%We compare the performance of the interpreters when running on SICStus and SWI Prolog, and obtained some surprising results.
Our ultimate goal is to assess which interpreter design in Prolog is the most efficient
as we intend to apply these results to a more complex language.
However, we believe the analysis in this paper to be of more general interest.

%This is a work-in-progress and other ideas to express the control flow are welcome.

\end{abstract}
\section{Introduction}
%\ds{Make sure we write bytecode in a uniformly}
Writing simple language interpreters in Prolog is pretty straightforward. 
Definite clause grammars (DCGs) enable parsing of the program,
and interpretation of the resulting abstract syntax tree (AST)
can be expressed in an idiomatic, recursive way:
%The
%data structures and language semantics are a natural match to the evaluation of
%programs, in particular if those are represented as trees.
Selecting which predicate to
execute in order to evaluate a part of a program is done by unifying the part of the program to be executed next
with the set of rules in Prolog's database that implement the language semantics.
Subsequent execution steps can be chosen by
using logic variables that are bound to substructures of the matched node.

Although this approach to interpreter construction is a natural match to
Prolog, the question remains if it is the most efficient way to implement the
instruction dispatching logic. % for any language
%implemented in Prolog.
In particular, we have developed such an interpreter \cite{DBLP:journals/sttt/LeuschelB08} for the entire B language~\cite{Abrial:BBook}
and want to evaluate the potential for improving its performance,
by using alternate implementation techniques.

Interpreters implemented in imperative languages, especially low-level
languages, often make use of alternative techniques to implement the
dispatching logic, taking advantage of available data structures and programming
paradigms. %that might be available in higher-level languages.
% of a language that all aim to reduce the overhead associated to deciding
% which instruction to execute next. E.g. by traversing the tree representation
% of the program in order to find the next node to be executed.

In this article, we explore if some of these techniques can be implemented in
Prolog or applied in interaction with a Prolog runtime with the goal to assess whether the instruction
dispatching for language interpreters can be made faster while keeping the language
semantics in Prolog.
In order to examine the performance of different dispatching models in Prolog,
we have defined a simple imperative language named \acol, which is briefly
described in \cref{language}. For \acol, we have created several implementations
described in \cref{implementations}, that use different paradigms for the
dispatching logic.
In \cref{evaluation}, we evaluate our approach on a set of benchmarks written in
\acol, executing the interpreters both on SICStus~\cite{carlsson2012sicstus} and SWI-Prolog~\cite{wielemaker_schrijvers_triska_lager_2012}.
Finally, we give our conclusions in \cref{sec:conclusions}.

\section{A Simple Language} \label{language}

As a means to evaluate different interpreter designs, % described in
%\cref{implementations},
we have defined a very simple and limited language
named \acol\footnote{\acol~is \emph{not} a backronym for \acol~is a computable
language.}. % but means "leading to the lake" in mapudungun
%that offers only a very limited set of features.

\acol~is an imperative language consisting of three kinds of statements:
while-loops, if-then-else statements and variable assignments. The only
supported value type is integer.
Furthermore, \acol~offers a few arithmetic operators (addition, subtraction,
multiplication and modulo), comparisons (less than (or equal to), greater than (or
equal to) and equals), as well as a boolean \texttt{not} operator.

% We would like to emphasise that \acol~is only intended to be used as a way to
% write small programs to be run on the different interpreters. Its set of
% features is very limited because it would have been very cumbersome to
% implement a more complex language in multiple ways. \acol~exclusively serves as
% a prototype to get an idea how well the different interpreter designs perform.

A simple \acol~program implementing a power function 
is shown in \cref{lst:example}.

\begin{figure}[t]
\begin{custom}
# the initial environment (i.e. input): base = 2, exponent = 5

# the program
val = 1;
while exponent > 0 {
    val = val * base;
    exponent = exponent - 1;
}
\end{custom}
\caption{An \acol{} program implementing a power function}
\label{lst:example}
\end{figure}

\section{Interpreter Implementations} \label{implementations}

There are many ways to implement \acol, in C as well as in Prolog.
Considering several interpreter implementation techniques, in this section,
we will describe possible designs of interpreters and the closely related representations
of the \acol~programs in Prolog. The interpreters are based on either traversing the
abstract syntax tree representation of a program or on compiling the program to
bytecode first and evaluating this more compact representation instead.

We opted to implement stack-based interpreters as their design tends to be simpler.
The alternative -- register-based virtual machines -- usually are faster~\cite{shi2008virtual}
and allow more advanced techniques such as register allocation optimisation
in order to reduce the amount of load and store instructions.
Yet, this endeavour would be far more involved and
could be considered if this prototype already shows proper speed-ups.

All interpreters share the same implementation of the language semantics
exposed by an object-space API~\cite{pypy:object_space}.
The objects space contains the code that creates integer objects,
performs arithmetic operations, compares values,
and manages the environment.
In order to keep the
implementations simple and compatible, all interpreters that we present
call into this same object space.
Nonetheless, the interpreters differ very much in the representation of the
program and, hence, in the process of dispatching.
The full code of all interpreters, benchmark scripts and results
can be found at:

\begin{center}
    \url{https://github.com/pkoerner/prolog-interpreters}
\end{center}

\begin{figure}[t]
  \centering
  \includegraphics[width=\linewidth]{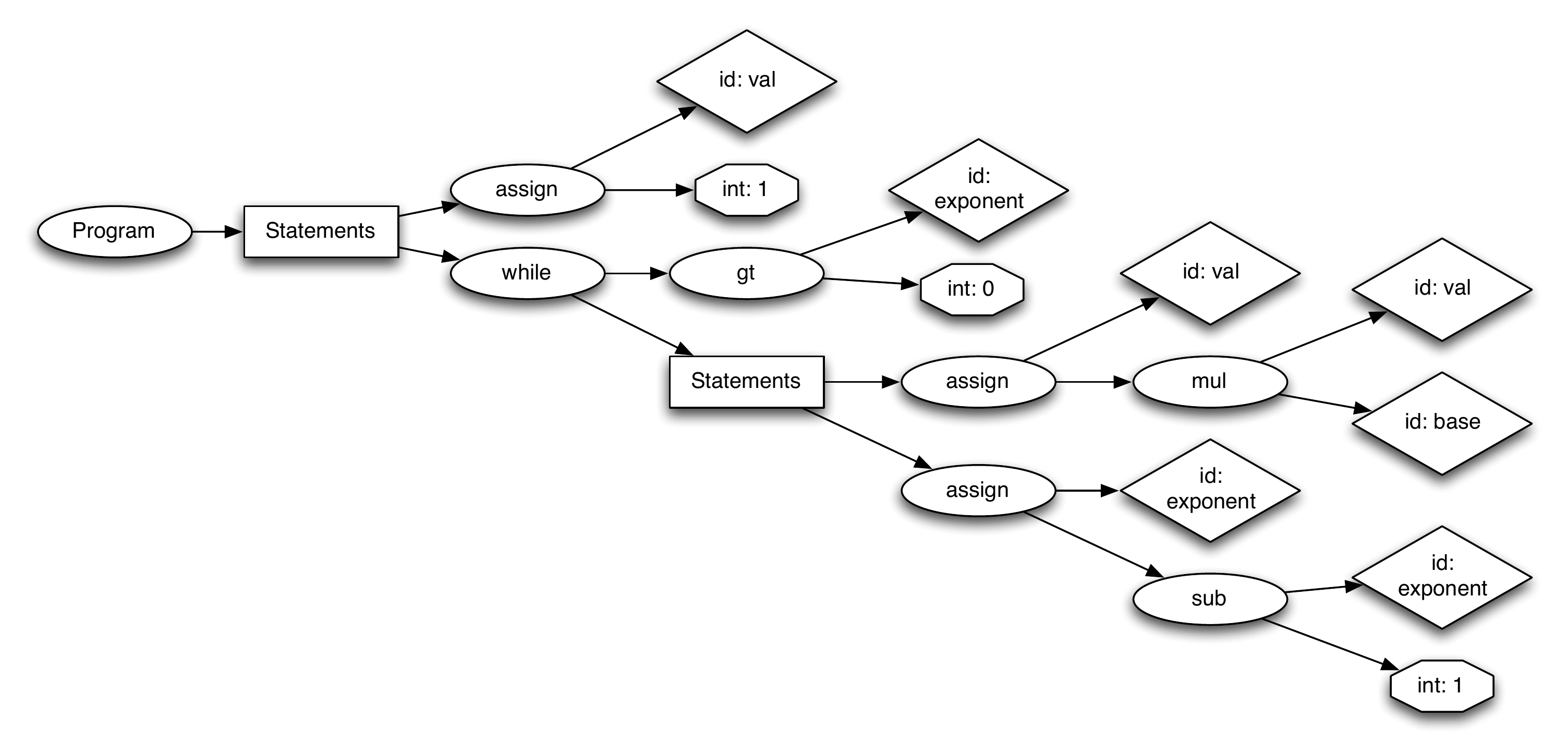}
  \caption{AST}
  \label{lst:example-ast}
\end{figure}

In order to discuss the differences, we will translate the small example program
shown in \cref{lst:example} into the different representations and show an
excerpt of the interpretation logic for each paradigm.
In \cref{lst:example-ast}, the AST for the example program is depicted.

\subsection{AST Interpreter}%
\label{sec:ast}

The most idiomatic way to implement an interpreter in Prolog is in form of an
AST-interpreter since it synergises very well with its execution model.

The data structure used for this interpreter is the tree representation of the
program as generated by the parser. In Prolog,
the AST can be represented as a single term as shown in \cref{lst:example-term}.
The program itself is a Prolog list of statements. However, every statement is
represented as its own sub-tree. Block statements, i.e. the body of \texttt{if} and
\texttt{while} instructions, will contain a list of statements themselves.

\begin{figure}[t]
\begin{custom}
[assign(id(val), int(1)),
 while(gt(id(exponent), int(0)),
       [assign(id(val), mul(id(val), id(base))),
        assign(id(exponent), sub(id(exponent), int(1)))])]
\end{custom}
\caption{Prolog representation of the AST}
\label{lst:example-term}
\end{figure}

\begin{figure}[t]
\begin{custom}
ast_int([], Env, _Objspace, Env).
ast_int([H|T], EnvIn, Objspace, EnvOut) :-
    ast_int(H, EnvIn, Objspace, Env), ast_int(T, Env, Objspace, EnvOut).
ast_int(if(Cond, Then, Else), EnvIn, Objspace, EnvOut) :-
    eval(Cond, EnvIn, Objspace, X),
    (X == true -> ast_int(Then, EnvIn, Objspace, EnvOut)
               ;  ast_int(Else, EnvIn, Objspace, EnvOut)).
ast_int(assign(id(Var), Expr), EnvIn, Objspace, EnvOut) :-
    eval(Expr, EnvIn, Objspace, Res), Objspace:store(EnvIn, Var, Res, EnvOut).
ast_int(while(Cond, Instr, _Invariant, _Variant), EnvIn, Objspace, EnvOut) :-
    ast_while(Cond, Instr, EnvIn, Objspace, EnvOut).
\end{custom}
\caption{Dispatching in a Prolog AST interpreter}
\label{lst:ast-interp-example}
\end{figure}

The AST interpreter will examine the first element of the list, execute this
statement and continue with the rest of the list, as can be seen in
\cref{lst:ast-interp-example}. Every sub-tree encountered this way is evaluated
recursively.

Choosing the implementation for each node in the tree is done by unifying the
current root node with the set of evaluation rules. This approach benefits from
the first argument indexing~\cite{Warren:1983uu} optimisation done by most
Prolog systems.

\subsection{Bytecode Interpreters} \label{bytecode}

We have defined a simple set of bytecodes, described below, as a compilation target for
\acol~programs. Based on these instructions we will introduce a series of
bytecode-interpreters that explore different implementation approaches in
Prolog and C.

As many bytecode interpreters for other languages, ours are \textit{stack-based}.
Some opcodes may create or load objects and store them on the evaluation stack,
e.g. \texttt{push} or \texttt{load}. Yet others may in turn consume objects from
the stack and create a new one in return, e.g. \texttt{add}.
Lastly, a single opcode is used to manipulate the environment, i.e.
\texttt{assign}.
An exhaustive list is shown in \cref{tbl:bytecode}.

\begin{table}[t]
\caption{A bytecode for the described language}
\label{tbl:bytecode}
\resizebox{\linewidth}{!}{
\begin{tabular}{c c c c}\toprule
\# & Name           & Arguments                   & Semantics\\\midrule
10 & jump           & 4 bytes encoded PC          & jumps to new PC\\ 
11 & jump-if-false  & 4 bytes encoded PC          & jumps to new PC if top element is falsey\\
12 & jump-if-true   & 4 bytes encoded PC          & jumps to new PC if top element is truthy\\
20 & push1          & 1 byte encoded integer      & push the argument on the stack\\
21 & push4          & 4 bytes encoded integer     & push the argument on the stack\\
40 & load           & 4 bytes encoded variable ID & push variable on the stack\\
45 & assign         & 4 bytes encoded variable ID & store top of the stack in variable\\
197 & mod           & -                           & pop operands, push result of operation\\
198 & mul           & -                           & pop operands, push result of operation\\
199 & sub           & -                           & pop operands, push result of operation\\
200 & add           & -                           & pop operands, push result of operation\\
240 & not           & -                           & pop operand, push negation\\
251 & eq            & -                           & pop operands, push result of comparison\\
252 & le            & -                           & pop operands, push result of comparison\\
253 & lt            & -                           & pop operands, push result of comparison\\
254 & ge            & -                           & pop operands, push result of comparison\\
255 & gt            & -                           & pop operands, push result of comparison\\\bottomrule
\end{tabular}}
\end{table}

\subsubsection{Imperative Bytecode Interpreter} \label{c-bytecode}

Usually, bytecode interpreters are written in imperative languages, that are
rather low-level, e.g. C, that allow more control about how objects are laid
out in memory and provide fine-grained control over the flow of execution.

To introduce the concept of a bytecode interpreter, we present an
implementation of \acol~beyond Prolog, that is purely written in C.

The bytecode is stored as a block of memory, that can be interpreted as an
array of bytes. The index of this array that should be interpreted next is
called the program counter. After that opcode is executed, the program counter
is incremented by one, plus the size of its arguments. However, it may be set to
an arbitrary index by opcodes implementing jumps. Integer arguments are encoded in reverse byte order.
%An example for a bytecode based on the program above is shown in
%\cref{lst:example-bytecode-c}.

%\begin{figure}[t]
%\begin{custom}
%unsigned char bc[] = {20, 1,           // push integer 1 on the stack
%                      45, 2, 0, 0, 0,  // store it in variable at index 2
%                                       // (i.e. val)
%                      40, 1, 0, 0, 0,  // load the variable at index 1
%                                       // (i.e. exponent)
%                      20, 0,           // push 0
%                      255,             // greater than
%                      11, 54, 0, 0, 0, // jump behind loop
%                                       // if condition is falsey
%                      40, 2, 0, 0, 0,  // load val
%                      40, 0, 0, 0, 0,  // load base
%                      198,             // mul
%                      45, 2, 0, 0, 0,  // store val
%                      40, 1, 0, 0, 0,  // load exponent
%                      20, 1,           // push 1
%                      199,             // sub
%                      45, 1, 0, 0, 0,  // store exponent
%                      10, 7, 0, 0, 0,  // jump to beginning of loop
%                      0}               // terminate
%\end{custom}
%\caption{Example bytecode in C}
%\label{lst:example-bytecode-c}
%\end{figure}

\begin{figure}[t]
\begin{minipage}[t]{0.95\textwidth}
\begingroup
\parfillskip=0pt
\begin{minipage}[c]{0.47\textwidth}
%\begin{figure}
\begin{custom}
while (pc < bc_len) {
    unsigned char *arg = bc + pc + 1;
    switch (bc[pc]) {
        case JUMP:
            pc = decode_arg4(arg); break;
        case LOAD:
            index = decode_arg4(arg);
            push(stack, env[index]);
            pc += 5; break;
        case ASSIGN:
            env[arg] = pop(stack);
            pc += 5; break;
        case ADD:
            b = pop(stack);
            a = pop(stack);
            push(stack, add(a, b));
            pc++; break;
        // ... many further cases
    }
}
\end{custom}
\captionof{figure}{Dispatching logic in C}
\label{lst:dispatching-c}
%\end{figure}
\end{minipage}%
\hfill
\begin{minipage}[c]{0.47\textwidth}
%\begin{figure}
\begin{custom}
while (pc < bc_len) {
    unsigned char *arg = bc + pc + 1;
    switch (bc[pc]) {
        case JUMP:
            pc = decode_arg4(arg); break;
        case LOAD:
            index = decode_arg4(arg)
            push(stack, env[index]);
            pc += 5; break;
        case ASSIGN:
            index = decode_arg4(arg);
            PL_put_term(env[index], pop(s));
            pc += 5; break;
        case ADD:
            arg1 = PL_new_term_refs(3);
            arg2 = arg1 + 1;
            var = arg1 + 2;
            PL_put_term(arg2, pop(s));
            PL_put_term(arg1, pop(s));
            PL_call_predicate(NULL,
                              PL_Q_NORMAL,
                              predicate_add,
                              arg1);
            push(s, var);
            pc++; break;
        // ...  many further cases
    }
}
\end{custom}
\captionof{figure}{Dispatching logic using SWI-Prolog's C-Interface}
\vspace{0.157in} % PI / 20 should be fine
\label{lst:dispatching-swi-c}
%\end{figure}
\end{minipage}%
\par\endgroup
\end{minipage}
\end{figure}

The dispatching logic is implemented as a \texttt{switch}-statement that is
contained in a loop. An excerpt of the implementation of our bytecode-interpreter in C
is shown in \cref{lst:dispatching-c}.
Every \texttt{case} block contains an implementation of that specific opcode.
After the opcode is executed, the program counter is advanced or reset and the
next iteration of the main loop is commenced.

\subsubsection{C-Interfaces} \label{sec:c-interface}

We made the digression into an interpreter written in C not only to present the
concept of bytecode interpreters. Instead, we can utilise the same dispatching
logic, but instead of calling an object space that is implemented in C, we can
use the C interfaces provided by the Prolog runtimes we consider (SICStus and
SWI-Prolog) to call arbitrary Prolog predicates. This way, we can query
the aforementioned object space that contains the semantics of \acol, but is
implemented in Prolog. An excerpt when using the C interface of SWI-Prolog is
shown in \cref{lst:dispatching-swi-c}.

For the C-interface, we re-use the linear bytecode from the Prolog interpreter above.
The list of bytecodes is passed to C, which allocates a C array,
iterates over the list and copies the instructions into the array.
Then, the main loop dispatches in C, but the objects on the evaluation stack
are created and the operations are executed by Prolog predicates.

\subsubsection{Prolog Facts} \label{sec:facts}

The main issue with bytecode interpreters in Prolog is the efficient implementation of
jumps to other parts of the bytecode. With an interpreter in C, all we have to
do is to re-assign the program counter variable. Prolog, however, does not
offer arrays with constant-time indexing\footnote{While, again, interoperability with C allows embedding of such data structures,
standard library predicates usually only offer logarithmic access.}.

The idiomatic way to simulate an array would be to use a Prolog list, but on
this data structure we can perform lookups only in $\mathcal{O}(n)$.
Yet, there are other representations of the program that allow jumping to
another position faster.

One way to express such a lookup in $\mathcal{O}(1)$ is to transform the
bytecode into Prolog terms
\texttt{bytecode(ProgramCounter, Instruction, Arguments)}.
Those terms are written into a seperate Prolog module that is loaded afterwards.
The first argument indexing optimisation then allows lookups in
constant time.

In contrast to an interpreter written in C, it does not perform well to encode
integer arguments into reverse byte-order arguments. Instead, we use the Prolog
primitives, i.e. integers for values and atoms for variable identifiers.

\begin{figure}[t]
\begin{custom}
bytecode(0, 20, 1).         % push integer 1 on the stack
bytecode(2, 45, val).       % pop value from stack, store in val
bytecode(7, 40, exponent).  % push value of exponent
bytecode(12, 20, 0).        % push constant 0
bytecode(14, 255, []).      % greater-than comparison
bytecode(15, 11, 54).       % jump-if-false to location 55 (exit loop)
bytecode(20, 40, val).      % push value of val
bytecode(25, 40, base).     % push value of base
bytecode(30, 198, []).      % multiplication of arguments on the stack
bytecode(31, 45, val).      % store result in val
bytecode(36, 40, exponent). % load exponent
bytecode(41, 20, 1).        % push constant 1
bytecode(43, 199, []).      % subtract arguments on stack
bytecode(44, 45, exponent). % store result in exponent
bytecode(49, 10, 7).        % jump to beginning of loop
bytecode(54, 0, []).        % terminate instruction
\end{custom}
\caption{Bytecode as Prolog facts}
\label{lst:example-facts}
\end{figure}

%unsigned char bc[] = {20, 1,           // push integer 1 on the stack
%                      45, 2, 0, 0, 0,  // store it in variable at index 2
%                                       // (i.e. val)
%                      40, 1, 0, 0, 0,  // load the variable at index 1
%                                       // (i.e. exponent)
%                      20, 0,           // push 0
%                      255,             // greater than
%                      11, 54, 0, 0, 0, // jump behind loop
%                                       // if condition is falsey
%                      40, 2, 0, 0, 0,  // load val
%                      40, 0, 0, 0, 0,  // load base
%                      198,             // mul
%                      45, 2, 0, 0, 0,  // store val
%                      40, 1, 0, 0, 0,  // load exponent
%                      20, 1,           // push 1
%                      199,             // sub
%                      45, 1, 0, 0, 0,  // store exponent
%                      10, 7, 0, 0, 0,  // jump to beginning of loop
%                      0}               // terminate
%\end{custom}

\Cref{lst:example-facts} shows a module that is generated from the bytecode.
The interpreter fetches the instruction located at the current program
counter, executes it and increments the program counter accordingly.
This is repeated until it encounters a special zero instruction denoting
the end of the bytecode -- here at location 54.

The dispatching mechanism is shown in \cref{lst:dispatching-facts}.
Similar to an interpreter in C, every opcode has an implementation in Prolog
that calls into the object space. Any rule of \texttt{fact\textunderscore{}int}
is equivalent to a \texttt{case} statement in C.

\begin{figure}[t]
\begin{custom}
fact_int(PC, Objspace, Env, Stack, REnv) :-
    generated:bc(PC, Instr, Args), % fetch the instruction
    fact_int(Instr, Args, PC, Stack, Env, Objspace, REnv).
fact_int(200, _Args, PC, [Y, X|Stack], Env, Objspace, REnv) :-
    Objspace:add(X, Y, Res), NewPC is PC + 1,
    fact_int(NewPC, Objspace, Env, [Res|Stack], REnv).
% fact_int also has implementations of all the other bytecodes...
\end{custom}
\caption{Dispatching in the facts-based interpreter}
\label{lst:dispatching-facts}
\end{figure}

\subsubsection{Sub-Bytecodes} \label{sec:subbytecode}

Another design is based on the idea that a program is executed
\textit{block-wise}, i.e. a series of instructions that is guarenteed to be
executed in this specific order. This is very simple since \acol~does not include a
goto-statement that allows arbitrary jumps. From a programmer's point of view,
blocks are the bodies of while-loops or those of if-then-else statements.

Instead of linearising the entire bytecode, only a block is linearised at
once. In order to deal with blocks that are contained by another block (e.g.
nested loops), two special opcodes are added. They are used to suspend the execution of the
current block and look up the \textit{sub-bytecodes} of the contained blocks that
are referenced via its arguments. After those sub-bytecodes are executed, the
execution of the previous
bytecode is resumed.

The special if-opcode references the blocks of the corresponding then- and else-
branches. After the condition is evaluated, only the required block is looked up
and executed. The other special opcode for while-loops references the bytecode
of the condition that is expected to leave true or false on the stack, as well
as the body of the loop. The blocks corresponding to condition and body are
evaluated in turn until the condition does not hold any more, so the execution
of its parent block can continue.
Similar to the facts in the interpreter above, the sub-bytecodes are asserted
into their own module to allow fast lookups.

\begin{figure}[t]
\begin{custom}
[20, 1, 45, val, % val = 1
 2, 0, 1]        % while (condition encoded in sub-bytecode 0,
                 %        body encoded in sub-bytecode 1)

% Sub-bytecodes
sbc(0, [40, exponent, 20, 0, 255]).
sbc(1, [40, val, 40, base, 198, 45, val, 40, exponent, 20, 1, 199]).
\end{custom}
\caption{Bytecode with sub-bytecodes}
\label{lst:example-subbytecodes}
\end{figure}

\Cref{lst:example-subbytecodes} shows an
example that includes the special opcode for the while-statement, and
\cref{lst:dispatching-subbytecodes} shows an excerpt of the dispatching logic
used for this interpreter.
The recursion in \texttt{bc\textunderscore{}int2} will update the
bytecode-list with its tail instead of manipulating a program counter.
Hence, in this implementation, the interpreter can only move forward inside of
a block. If it is required to move backwards in the program, it is only possible
to re-start at the beginning of a block.

\begin{figure}[t]
\begin{custom}
bc_int([], Env, Stack, _Objspace, Env, Stack).
bc_int([H|R], Env, Stack, Objspace, REnv, RStack) :-
    bc_int2(H,R, Env, Stack, Objspace, REnv, RStack).
% special bytecodes for evaluating blocks of an if-statement
bc_int2(1, [T, E|R], Env, [Cond|Stack], Objspace, REnv, RStack) :-
    (Cond == true -> subbytecodes:sbc(T, Then),
                     h_bc_int(Then, [], Env, Objspace, TEnv)
                  ;  subbytecodes:sbc(E, Else),
                     h_bc_int(Else, [], Env, Objspace, TEnv)),!,
    bc_int(R, TEnv, Stack, Objspace, REnv, RStack).
% special bytecodes for evaluating blocks of a while-loop
bc_int2(2, [C, I|R], Env, Stack, Objspace, REnv, RStack) :-
    subbytecodes:sbc(C, Cond),
    bc_int(Cond, Env, [], Objspace, Env, [Res]),
    (Res == true -> subbytecodes:sbc(I, Instr),
                    h_bc_int(Instr, [], Env, Objspace, T),!,
                    bc_int2(2, [C, I|R], T, Stack, Objspace, REnv, RStack)
                 ;  !, bc_int(R, Env, Stack, Objspace, REnv, RStack)).

bc_int2(200, R, Env, [Y, X|Stack], Objspace, REnv, RStack) :-
    Objspace:add(X, Y, Res),!,
    bc_int(R, Env, [Res|Stack], Objspace, REnv, RStack).
% bc_int2 also has implementations of all the other bytecodes...
\end{custom}
\caption{Dispatching on bytecodes with sub-bytecodes}
\label{lst:dispatching-subbytecodes}
\end{figure}

\subsection{Rational Trees}\label{sec:rt}

Based on \cite{Carro:2004vh}, we have created implementations of an AST- and a
bytecode-interpreter for \acol~that use the idea of rational trees to represent
the program being evaluated. This technique aims to improve the performance of
jumps by using recursive data structures containing references to the following
instructions.

\subsubsection{AST-Interpreter with Rational Trees} \label{sec:rt-ast}

Since \acol~does not include a concept of arbitrary jumps as used in
\cite{Carro:2004vh}, it is not possible to achieve the speed-up described
in the referenced paper. However, we can make use of the basic idea for the
representation of programs: every statement has a pointer to its successor
statement. 

In our naive AST interpreter, a new Prolog stack frame is used for every level
of nested loops and if-statements. Instead of returning from each evaluation to the
predicate that dispatched to the sub-statement, we can make use of Prolog's
tail-recursion optimisation and continue with the next statements directly.

\begin{figure}[t]
\begin{custom}
assign(id(val), int(1),
  while(gt(id(exponent), int(0)),
    assign(id(val), mul(id(val), id(base)),
      assign(id(exponent), sub(id(exponent), int(1)),
        while(gt(id(exponent), int(0)),
         ...)))
    end))
\end{custom}
\caption{Rational tree representation}
\label{lst:example-rational-tree}
\end{figure}

For our example program, we generate an infinite data structure for the
while-loop depicted in \cref{lst:example-rational-tree}.
The concept of rational trees allows us to have the \texttt{while}-term
re-appearing in its own body, so it has not to be saved in a stack frame.

The last statement \texttt{end} is artificially added to indicate the end of
the program so that the interpreter may halt.

Then, the dispatching logic is still very similar to the naive AST
interpreter as shown in \cref{lst:dispatching-rational-tree}.

\begin{figure}[t]
\begin{custom}
rt_int(end, Env, _, Env) :- !.
rt_int(assign(id(Var), Expr, Next), Env, Objspace, REnv) :-
    eval(Expr, Env, Objspace, Res),
    Objspace:store(Env, Var, Res, EnvOut), !,
    rt_int(Next, EnvOut, Objspace, REnv).
rt_int(if(Cond, Then, Else), Env, Objspace, REnv) :-
    eval(Cond, Env, Objspace, V),
    (V == true -> !, rt_int(Then, Env, Objspace, REnv)
                ; !, rt_int(Else, Env, Objspace, REnv)).
rt_int(while(Cond, Instrs, Else), Env, Objspace, REnv) :-
    eval(Cond, Env, Objspace, V),
    (V == true -> !, rt_int(Instrs, Env, Objspace, REnv)
                ; !, rt_int(Else, Env, Objspace, REnv)).
\end{custom}
\caption{Dispatching in a rational tree interpreter}
\label{lst:dispatching-rational-tree}
\end{figure}

\subsubsection{Bytecode-Interpreter with Rational Trees}

In Prolog, rational trees can also be used for bytecodes. Jumps are removed
from that representation entirely. While-loops are unrolled into an infinite
amount of alternated bytecodes of the condition and if-statements that contain
the body of the loop in their then-branch and the next statement after the loop
in their else-branch. An example is shown in \cref{lst:example-bc-rational-tree}.

At first glance, it looks weird that the opcode integers are replaced by
human-readable descriptions.
However, functors are limited to atoms and, then, there is not much difference
between atoms that contain only a number or short readable names. We chose the
latter one because they are by far more comprehensible.

\begin{figure}[t]
\begin{custom}
push(1, assign(val,                                  % code before the loop
  load(exponent, push(0, gt(                         % condition (1)
    if(load(val, load(base, mul(store(val,           % while-body (1)
         load(exponent, push(1, sub(store(exponent,  % while-body (1)
           load(exponent, push(0, gt(                % condition (2)
             if(load(val, load(base(, ....))),       % while-body (2)
                end))))))))))))                      % end of while (2)
       end))))))                                     % end of while (1)
\end{custom}
\caption{Bytecode with rational trees}
\label{lst:example-bc-rational-tree}
\end{figure}

\begin{figure}[t]
\begin{custom}
rt_bc_int(end, Env, Stack, _Objspace, Env, Stack).
rt_bc_int(if(Then, Else), Env, [X|Stack], Objspace, REnv, RStack) :-
    (X == true -> !, rt_bc_int(Then, Env, Stack, Objspace, REnv, RStack)
                ; !, rt_bc_int(Else, Env, Stack, Objspace, REnv, RStack)).
rt_bc_int(push(Arg, Next), Env, Stack, Objspace, REnv, RStack) :-
    Objspace:create_integer(Arg, Val),!,
    rt_bc_int(Next, Env, [Val|Stack], Objspace, REnv, RStack).
rt_bc_int(load(Arg, Next), Env, Stack, Objspace, REnv, RStack) :-
    Objspace:lookup(Arg, Env, Val), !,
    rt_bc_int(Next, Env, [Val|Stack], Objspace, REnv, RStack).
rt_bc_int(add(Next), Env, [Y, X|Stack], Objspace, REnv, RStack) :-
    Objspace:add(X, Y, Res), !,
    rt_bc_int(Next, Env, [Res|Stack], Objspace, REnv, RStack).
% rt_bc_int implements all other opcodes as well...
\end{custom}
\caption{Dispatching in a bytecode interpreter with rational trees}
\label{lst:dispatching-bc-rational-tree}
\end{figure}

The dispatching is pretty similar to the AST interpreter that utilises
rational trees, as shown in \cref{lst:dispatching-bc-rational-tree}.
The main difference between those two interpreters is that this one uses a
simulated stack to evaluate terms instead of Prolog's call stack.

\section{Evaluation} \label{evaluation}

To compare the performance of the different interpreters for \acol, we selected
a set of different benchmarks. Because the language is very limited, it is hard
to design "real-world programs".
Yet, execution of any arbitrary program will give insight of the performance
of the dispatching logic.

In this section, we present those benchmarks and compare their results.
Each program was executed with every interpreter ten times. The runtime consists
only of the time spent in the interpreter. 
Compilation time is excluded, as it is not implemented efficiently and,
ultimately, not relevant.

The benchmarks were run on a machine that runs a linux with a 4.15.0-108-generic
64-bit kernel on an Intel i7-7700HQ CPU @ 2.80GHz.
No benchmarks ran in parallel.
Two Prolog implementations were considered: SICStus Prolog 4.6.0, a commercial
product, and SWI-Prolog 8.2.1, a free open-source implementation. All C code
was compiled by gcc 7.5.0 with the \texttt{-O3}-flag.

Since \acol~does not offer complex features, we expect that the dispatching
claims a bigger share of the runtime than the actual operations.

\subsection{Benchmarks}

\paragraph{Prime Tester}

The first benchmark is a naive prime tester. The program is depicted in
\cref {lst:benchmark-prime}. The environment was pre-initialised with
$\texttt{is\textunderscore{}prime} \coloneqq 1$,
$\texttt{start} \coloneqq 2$,
and $\texttt{V} \coloneqq \num{34265341}$.

\begin{figure}[t]
\begin{custom}
while (start < V) {
    if (V mod start == 0) {
        is_prime := 0;
    } else {
        is_prime := is_prime;
    }
    start := start + 1;
}
\end{custom}
\caption{Prime Tester Program}
\label{lst:benchmark-prime}
\end{figure}

\paragraph{Fibonacci}

Another benchmark is the calculation of the fibonacci sequence.
However, we expect that most of the execution time will consist of the addition
and subtraction of two large numbers and that the interpreter overhead itself is
rather small. Therefore, a second version that calculates the sequence modulo
$\num{1000000}$ is included.

\begin{figure}[t]
\begin{custom}
i := 1;                     i := 1;
while i < n {               while i < n {
    b := b + a;                 b := b + a mod 1000000;
    a := b - a;                 a := b - a mod 1000000;
    i := i + 1;                 i := i + 1;
}                           }
\end{custom}
\caption{Fibonacci Programs}
\label{lst:benchmark-fib}
\end{figure}

Again, the environment is pre-initialised, in this case with
$\texttt{a} \coloneqq 0$,
$\texttt{b} \coloneqq 1$ and
$\texttt{n} \coloneqq \num{400000}$. To ensure a significant runtime for the
second version, the input is modified so it calculates a longer sequence, i.e.
$\texttt{n} \coloneqq \num{10000000}$.

\paragraph{Generated ASTs}

Lastly, some programs were generated pseudo-randomly. Such a generated AST
consists of 20 to 50 statements that are uniformly chosen from while-loops,
if-statements and assignments. The body of a loop and both branches of
if-statements also consist of 20 to 50 statements. However, if the nesting
exceeds a certain depth, only assignments are generated for this block.

In order to guarentee termination, while-loops are always executed 20 times.
An assignment is artifically inserted before the loop that resets a loop
counter, as well as another assignment that increments this variable at the
beginning of the loop.

For assignments and if-conditions, a small subtree is generated. The generator
chooses uniformly between five predetermined identifiers, constants ranging
from -1 to 3, as well as additions and subtractions. If-conditions have to
include exactly one comparison operator.

The generator does include neither multiplications, because they caused very
large integers that slowed down the Prolog execution time significantly, nor
modulo operations, to avoid division by zero errors.

Three different benchmarks were generated using arbitrary seeds. Their purpose
is to complement the other three handwritten benchmarks, which are rather
small and might benefit from caching of the entire AST.

\subsection{Results}

\begin{table}
\centering
\caption{Mean runtimes in seconds including the 0.95 confidence interval.
The value in parentheses describes the normalised runtime
(on the basis of the AST interpreter). The fastest runtimes per benchmark and interpreter
are highlighted.}
\label{tbl:runtimes}

\setlength\tabcolsep{.4em}
\begin{tabular}{l  l rcr rcr} \toprule
\multicolumn{1}{c  }{Benchmark}  & \multicolumn{1}{c }{Interpreter}& \multicolumn{3}{c }{SICStus}& \multicolumn{3}{c}{SWI-Prolog}\\  \midrule 
\multirow{6}{*}{Prime Tester}       & AST                   &       54.53 &      {$\pm$} &      {\tiny 0.49 (1.00)} &       242.84 &      {$\pm$} &       {\tiny 3.96 (1.00)} \\
                                    & Sub-Bytecodes         &       78.03 &      {$\pm$} &      {\tiny 0.59 (1.43)} &       316.44 &      {$\pm$} &       {\tiny 1.89 (1.30)} \\
                                    & Facts                 &       73.50 &      {$\pm$} &      {\tiny 9.39 (1.35)} &       330.89 &      {$\pm$} &       {\tiny 2.51 (1.36)} \\
                                    & C-Interface           &      119.94 &      {$\pm$} &      {\tiny 5.13 (2.20)} &\best   54.83 &\best {$\pm$} &\best  {\tiny 0.98 (0.23)} \\
                                    & AST w/ Rational Trees &\best  54.26 &\best {$\pm$} &\best {\tiny 0.57 (1.00)} &       229.65 &      {$\pm$} &       {\tiny 2.28 (0.95)} \\
                                    & BC w/ Rational Trees  &       70.65 &      {$\pm$} &      {\tiny 1.19 (1.30)} &       261.39 &      {$\pm$} &       {\tiny 5.58 (1.08)} \\ \midrule
\multirow{6}{*}{Fibonacci}          & AST                   &        9.86 &      {$\pm$} &      {\tiny 0.05 (1.00)} &         5.54 &      {$\pm$} &       {\tiny 0.09 (1.00)} \\
                                    & Sub-Bytecodes         &       10.16 &      {$\pm$} &      {\tiny 0.13 (1.03)} &         6.36 &      {$\pm$} &       {\tiny 0.15 (1.15)} \\
                                    & Facts                 &       10.00 &      {$\pm$} &      {\tiny 0.07 (1.01)} &         6.49 &      {$\pm$} &       {\tiny 0.09 (1.17)} \\
                                    & C-Interface           &       10.16 &      {$\pm$} &      {\tiny 0.09 (1.03)} &\best    2.58 &\best {$\pm$} &\best  {\tiny 0.06 (0.47)} \\
                                    & AST w/ Rational Trees &        9.84 &      {$\pm$} &      {\tiny 0.19 (1.00)} &         5.39 &      {$\pm$} &       {\tiny 0.09 (0.97)} \\
                                    & BC w/ Rational Trees  &\best   9.83 &\best {$\pm$} &\best {\tiny 0.06 (1.00)} &         5.63 &      {$\pm$} &       {\tiny 0.10 (1.02)} \\ \midrule
\multirow{6}{*}{Fibonacci (Maxint)} & AST                   &\best  24.03 &\best {$\pm$} &\best {\tiny 0.30 (1.00)} &        96.90 &      {$\pm$} &       {\tiny 1.72 (1.00)} \\
                                    & Sub-Bytecodes         &       31.03 &      {$\pm$} &      {\tiny 0.26 (1.29)} &       122.86 &      {$\pm$} &       {\tiny 1.54 (1.27)} \\
                                    & Facts                 &       30.13 &      {$\pm$} &      {\tiny 0.48 (1.25)} &       131.38 &      {$\pm$} &       {\tiny 3.68 (1.36)} \\
                                    & C-Interface           &       42.23 &      {$\pm$} &      {\tiny 0.90 (1.76)} &\best   22.65 &\best {$\pm$} &\best  {\tiny 0.48 (0.23)} \\
                                    & AST w/ Rational Trees &       24.05 &      {$\pm$} &      {\tiny 0.17 (1.00)} &        94.36 &      {$\pm$} &       {\tiny 1.15 (0.97)} \\
                                    & BC w/ Rational Trees  &       28.63 &      {$\pm$} &      {\tiny 0.25 (1.19)} &       103.93 &      {$\pm$} &       {\tiny 0.70 (1.07)} \\ \midrule
\multirow{6}{*}{Generated}          & AST                   &       12.96 &      {$\pm$} &      {\tiny 0.16 (1.00)} &        60.64 &      {$\pm$} &       {\tiny 0.59 (1.00)} \\
                                    & Sub-Bytecodes         &       19.71 &      {$\pm$} &      {\tiny 0.70 (1.52)} &        69.26 &      {$\pm$} &       {\tiny 0.49 (1.14)} \\
                                    & Facts                 &       20.76 &      {$\pm$} &      {\tiny 0.33 (1.60)} &        73.89 &      {$\pm$} &       {\tiny 0.45 (1.22)} \\
                                    & C-Interface           &       24.49 &      {$\pm$} &      {\tiny 0.82 (1.89)} &\best   10.30 &\best {$\pm$} &\best  {\tiny 0.18 (0.17)} \\
                                    & AST w/ Rational Trees &\best  12.89 &\best {$\pm$} &\best {\tiny 0.14 (0.99)} &        58.35 &      {$\pm$} &       {\tiny 1.45 (0.96)} \\
                                    & BC w/ Rational Trees  &       16.90 &      {$\pm$} &      {\tiny 0.20 (1.30)} &        61.98 &      {$\pm$} &       {\tiny 1.16 (1.02)} \\ \midrule
\multirow{6}{*}{Generated2}         & AST                   &\best  19.01 &\best {$\pm$} &\best {\tiny 0.18 (1.00)} &        83.18 &      {$\pm$} &       {\tiny 0.80 (1.00)} \\
                                    & Sub-Bytecodes         &       29.11 &      {$\pm$} &      {\tiny 0.48 (1.53)} &       102.17 &      {$\pm$} &       {\tiny 1.48 (1.23)} \\
                                    & Facts                 &       30.78 &      {$\pm$} &      {\tiny 0.55 (1.62)} &       109.08 &      {$\pm$} &       {\tiny 2.35 (1.31)} \\
                                    & C-Interface           &       35.89 &      {$\pm$} &      {\tiny 0.92 (1.89)} &\best   15.07 &\best {$\pm$} &\best  {\tiny 0.24 (0.18)} \\
                                    & AST w/ Rational Trees &       19.10 &      {$\pm$} &      {\tiny 0.26 (1.00)} &        81.05 &      {$\pm$} &       {\tiny 0.39 (0.97)} \\
                                    & BC w/ Rational Trees  &       25.28 &      {$\pm$} &      {\tiny 2.96 (1.33)} &        90.44 &      {$\pm$} &       {\tiny 1.21 (1.09)} \\ \midrule
\multirow{6}{*}{Generated3}         & AST                   &\best  12.37 &\best {$\pm$} &\best {\tiny 0.22 (1.00)} &        55.52 &      {$\pm$} &       {\tiny 1.34 (1.00)} \\
                                    & Sub-Bytecodes         &       18.93 &      {$\pm$} &      {\tiny 0.20 (1.53)} &        66.00 &      {$\pm$} &       {\tiny 0.76 (1.19)} \\
                                    & Facts                 &       20.06 &      {$\pm$} &      {\tiny 0.64 (1.62)} &        70.49 &      {$\pm$} &       {\tiny 0.45 (1.27)} \\
                                    & C-Interface           &       23.51 &      {$\pm$} &      {\tiny 0.71 (1.90)} &\best    9.79 &\best {$\pm$} &\best  {\tiny 0.22 (0.18)} \\
                                    & AST w/ Rational Trees &       12.38 &      {$\pm$} &      {\tiny 0.30 (1.00)} &        52.96 &      {$\pm$} &       {\tiny 1.12 (0.95)} \\
                                    & BC w/ Rational Trees  &       16.10 &      {$\pm$} &      {\tiny 0.11 (1.30)} &        60.07 &      {$\pm$} &       {\tiny 0.97 (1.08)} \\ \bottomrule
\end{tabular}
\end{table}

\begin{figure}[t]
  \centering
  \includegraphics[width=\linewidth]{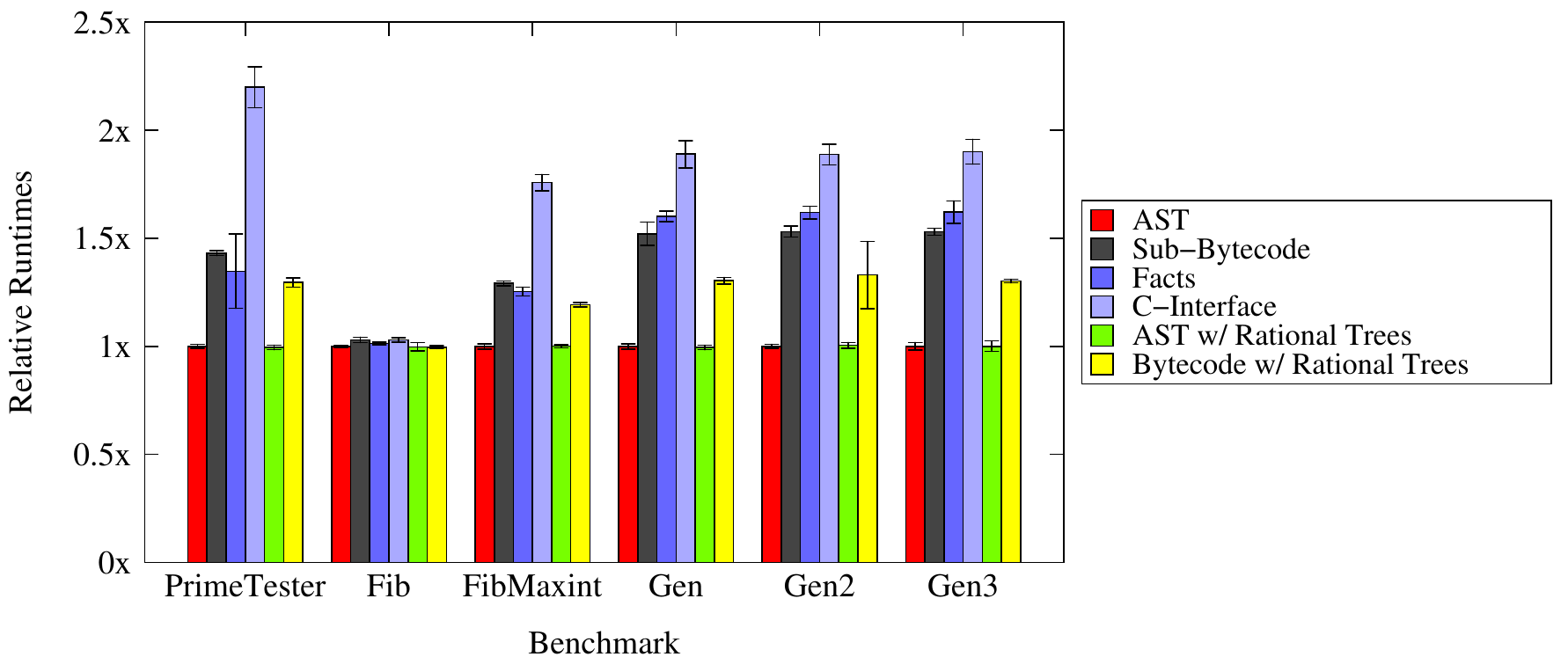}
  \caption{Relative runtimes in SICStus, normalised to the runtime of the AST interpreter}
  \label{fig:bargraphs-sics}
\end{figure}
\begin{figure}[t]
  \centering
  \includegraphics[width=\linewidth]{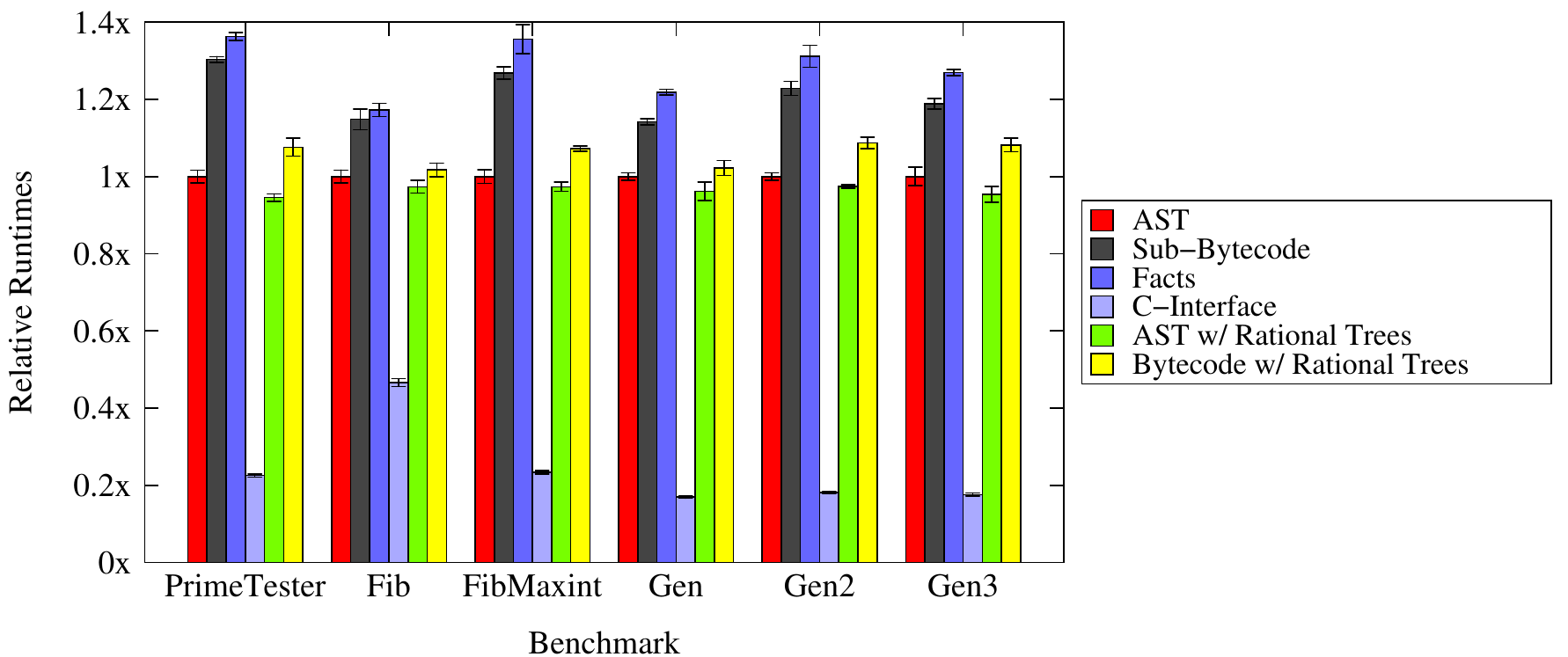}
  \caption{Relative runtimes in SWI-Prolog, normalised to the runtime of the AST interpreter}
  \label{fig:bargraphs-swi}
\end{figure}

The results of the benchmarks are shown in \cref{tbl:runtimes}.
The lines labelled \enquote{AST} refer to the implementation of the naive AST interpreter presented in \cref{sec:ast},
the ones with \enquote{Sub-Bytecodes}, \enquote{Facts} and \enquote{C-Interface} refer to the corresponding bytecode interpreters
discussed in \cref{sec:subbytecode}.
Finally, \enquote{AST-} and \enquote{BC w/ Rational Trees} are the AST and bytecode interpreters
based on rational trees presented in \cref{sec:rt}.
The mean value is determined by the geometric mean as proposed by
\cite{Fleming:1986:LSC:5666.5673}.
Though not listed, the interpreter purely written in C
is 2-3 orders of magnitudes faster\footnote{A fair comparison is not possible since the C interpreter does not support unbounded integer values.} and executes each benchmark in less than a second.

\Cref{fig:bargraphs-sics} shows the results specific for SICStus Prolog.
There is no discernible performance difference between
the naive AST interpreter and the ones utilising rational trees.
Independent of the benchmark, the bytecode interpreters based on sub-bytecodes
and on Prolog facts are slow in comparison.
One can observe a performance loss of about 25-35 \% for the small handwritten programs,
where we would expect caching effects to be the largest,
and around 50-60\% for the larger, generated programs.
With our initial version of the interpreter dispatching in C,
we reported an issue that was related with SICStus' FLI garbage collector.
Now, it usually requires twice as much time to execute the benchmarks
compared to the AST-based interpreters.

The results utilising SWI-Prolog are shown in \cref{fig:bargraphs-swi}.
Overall, they paint a similar picture to the results for SICStus.
However, the dispatching using SWI-Prolog's C-interface is very fast --
compared to the AST interpreter, it can achieve more than a 5$\times$ speed-up.

\section{Conclusions, Related and Future Work} %\ds{simplify title}
\label{sec:conclusions}

In this paper, we presented the language \acol~and multiple ways to implement
it as AST as well as bytecode interpreters. We designed several benchmarks in
order to evaluate their performance using different implementations of Prolog.

Our results suggest that if an interpreter is to be implemented in Prolog, the
implementation as an AST interpreter already is very performant.
It is simply not worth the hassle of writing and maintaining a bytecode compiler.
Furthermore, an AST interpreter does not
involve any additional compilation overhead 
as it can directly work on the data structure returned
by the parser.

However, SWI-Prolog's C interface performs very well.
Surprisinly, even on the Fibonacci example with unlimited integers,
where addition of unlimited integers is rather time-consuming,
it beats the run-time of the AST interpreter by a factor of two.
Additional work is required to determine whether these findings are applicable for more complex languages,
that would also facilitate the creation of more sensible benchmarks.

%one can utilise C to efficiently implement
%the dispatching and query Prolog predicates for the domain logic.

Rossi and Sivalingam explored dispatching techniques in C based bytecode
interpreters~\cite{Rossi:1996ue}, with the result that a less portable approach of composing the
code in memory before executing it yielded the best results. The techniques
discussed in that article could be used in combination with SWI-Prolog to
further improve the instruction dispatching performance in C.

An alternative for improving the execution time of a program, that was not discussed here, is
partial evaluation~\cite{Jones:peval}.
We intend to investigate the impact of offline partial evaluation when compiling a subset of the described interpreters
for our benchmarks.
%[A small discussion about partial evaluation \cite{Jones:peval}. We have used offline partial evaluation for Prolog \cite{LeuschelEtAl:TPLP03} to compile the AST interpreter for Fibonacci. The results are ...] \todo{Run the PE version for Fib and compare with results}

In the future, it would also be interesting to evaluate the effects
of different interpreter designs in other Prolog dialects,
especially those that are not based on the WAM~\cite{Ait-Kaci99warrensabstract}.
Examples include Ciao (WAM-based with powerful analysis),
BinProlog (specialised version of the WAM) and
Mercury (functional influences with many optimisations).

%However, \acol~is a very simple language. 
%Additional work is required to determine whether these findings are applicable for more complex languages.
%Furthermore, a richer language facilitates the creation of more benchmarks.

\bibliography{references}
\bibliographystyle{plain}
%\newpage
% http://tex.stackexchange.com/questions/103735/list-of-todos-todonotes-is-empty-with-llncs
%\setcounter{tocdepth}{1}
%\listoftodos
\end{document}